\documentclass[epj]{webofc}
\usepackage[varg]{txfonts}   
\wocname{EPJ Web of Conferences}
\woctitle{ICNFP 2017}

\usepackage{cancel} 
\usepackage{youngtab}
\usepackage{xfrac}
\usepackage{wrapfig}
\usepackage{adjustbox}
\begin{document}
\selectlanguage{english}
\title{Mass problem in the Standard Model}
%
%

\author{R. Martinez\inst{1}\fnsep\thanks{\email{remartinezm@unal.edu.co}} \and
        S. F. Mantilla\inst{1}\fnsep\thanks{\email{sfmantillas@unal.edu.co}} 
}

\institute{Departamento de F\'{i}sica, Universidad Nacional de Colombia, 
\\Ciudad Universitaria, Carrera 45 \# 26-85, Bogot\'{a} D.C., Colombia. 
}

\abstract{%
We propose a new $\mathrm{SU}(3)_C \otimes \mathrm{SU}(2)_{L}\otimes \mathrm{U}(1)_{Y}\otimes \mathrm{U}(1)_{X}$ gauge model which is non universal respect to the three fermion families of the Standard Model. We introduce additional one top-like quark, two bottom-like quarks and three right handed neutrinos in order to have an anomaly free theory. We also consider additional three right handed neutrinos which are singlets respect to the gauge symmetry of the model to implement see saw mechanism and give masses to the light neutrinos according to the neutrino oscillation phenomenology. In the context of this horizontal gauge symmetry we find mass ansatz for leptons and quarks. In particular, from the analysis of solar, atmospheric, reactor and accelerator neutrino oscillation experiments, we get the allow region for the Yukawa couplings for the charge and neutral lepton sectors according with the mass squared differences and mixing angles for the two neutrino hierarchy schemes, normal and inverted.
}
\maketitle
\section{Introduction}
\label{intro}
A large amount of high-energy phenomena have been understood in the context of the Standard Model (SM)\cite{SM}, nowadays one of the most successful frameworks in physics. Nevertheless, there are some observations which might be out of the scope of the SM such as the fermion mass hierarchy (FMH), the origin of the large differences among scales of fermion masses, from units of MeV to hundreds of GeV is not completely understood without using unpleasant fine-tunings. Moreover, according to neutrino detectors\cite{homestake}, the evidence of light neutrino masses from neutrino oscillations enlarges this issue by extending the mass scales until meV. However, in contrast with charged fermions, neutrino masses are known until their squared mass differences $\Delta m_{12}^{2}$ and $\Delta m_{\ell 3}$, where the last one determines the ordering, normal ordering (NO) if $\Delta m_{13}^{2}>0$ and inverted ordering (IO) if $\Delta m_{23}^{2}<0$\cite{nova,neutrinodata}. 

There are many models which propose different scenarios where FMH could be understood. One of the most important is the left-right model $\mathrm{SU(2)}_{L}\otimes \mathrm{SU(2)}_{R}\otimes \mathrm{U(1)}_{B-L}$\cite{fritzsch1978} from which the Fritzsch texture is obtained\cite{textures}. Such a scheme was extended to leptons and neutrinos by Fukugita, Tanimoto and Yanagida, who also implemented Majorana masses so as active neutrinos can acquire tiny masses, with the respective lepton violation processes produced by the existence of Majorana fermions\cite{fty1993}. 

Among different schemes, abelian extensions of the SM show a fertile scenario where some issues can be understood by implementing simple tools such as symmetry breaking or chiral anomaly cancellation. Their suitability has been shown in addressing quark masses\cite{somepheno}, dark matter\cite{DM-martinez-I,DM-jhep}, scalar potential stability\cite{DM-martinez-II} and lepton masses\cite{martinez1612}, where light fermions acquire masses through radiative corrections. However, by doing a slight modification in adding an additional doublet, a completely new scheme emerges with interesting texture matrices from which the FMH can be obtained naturally. 

The proceedings presents the particle content and the set of $\mathrm{U(1)}_{X}$ charges\cite{mantilla2017}. Thereafter, the mass matrices are shown from the Yukawa Lagrangians with their respective eigenvalues in which the FMH can be inferred. After that, the suitability of the model is checked by exploring the neutrino parameter space in order to fit neutrino oscillation data. Finally, some conclusions are outlined with a summary. 

\section{Particle content}
\label{particle-content}
The scheme proposed consists on extending the SM with a new nonuniversal abelian interaction $\mathrm{U(1)}_{X}$ together with a discrete symmetry $\mathbb{Z}_{2}$ which distinguish among families in such a way that fermion mass matrices can suggest the FMH without any kind of fine-tuning on Yukawa coupling constants. Moreover, because of the new gauge boson $Z'$, an extra singlet Higgs field $\chi$ is needed to break $\mathrm{U(1)}_{X}$ with its vacuum expectation value (VEV). Additionally, there are three Higgs doublets so as every fermion gets massive, reducing the need of radiative corrections to the minimum. There is also a scalar field $\sigma$ without VEV, whose $U(1)_{X}$ charge is the same of $\chi$ but with the opposite parity $\mathbb{Z}_{2}$ available to do radiative corrections when they are deserved. Furthermore, the scalar sector shows the vacuum hierarchy (VH) $v_{\chi}>v_{1}>v_{2}>v_{3}$, required to get the suited masses. $v_{\chi}$, $v_{1}$, $v_{2}$ and $v_{3}$ are at units of TeV, hundreds of GeV, units of GeV and hundreds of MeV, respectively.

On the other hand, the set of charges $\mathrm{U(1)}_{X}$ of the fermion sector is strongly constraint by the cancellation of chiral anomalies\cite{somepheno}. Thus, in order to obtain nonuniversal charges and also cancel any chiral anomaly in the model, exotic quarks $\mathcal{T},\mathcal{J}^{1,2}$ and leptons $\mathcal{E}^{1,2}$ have been included, together with right-handed neutrinos $\nu_{R}^{e,\mu,\tau}$ and Majorana fermions $\mathcal{N}_{R}^{1,2,3}$ so as inverse seesaw mechanism (ISS) is available to get small neutrino masses\cite{inverseseesaw}. All of them acquire mass through $\chi$ except $\mathcal{N}^{1,2,3}$ which have their own Majorana mass $M_{\mathcal{N}}$. Additionally, the $\mathbb{Z}_{2}$ is in charge to make distinguishable the fermions with the same $\mathrm{U(1)}_{X}$ charges. It is outlined in the condensed notation $X^{\pm}$ in tab. \ref{tab:Particle-content} of the particle content. 

\begin{table*}
\centering
\begin{adjustbox}{width=10cm,center}
\begin{tabular}{cc|cc|cc}
\hline\hline
Bosons	&	$X^{\pm}$	&	Quarks	&	$X^{\pm}$	&	Leptons	&	$X^{\pm}$	\\ \hline 
\multicolumn{2}{c}{Scalar Doublets}	&
\multicolumn{4}{c}{SM Fermionic Doublets}	\\ \hline\hline
$\Phi_{1}=\left(\begin{array}{c}
\phi_{1}^{+} \\ \frac{h_{1}+v_{1}+i\eta_{1}}{\sqrt{2}}	
\end{array}\right)$	&	${+2}/{3}^{+}$	&
$q^{1}_{L}=\left(\begin{array}{c}u^{1}	\\ d^{1} \end{array}\right)_{L}$
	&	${+1}/{3}^{+}$	&	
$\ell^{e}_{L}=\left(\begin{array}{c}\nu^{e} \\ e^{e} \end{array}\right)_{L}$
	&	$0^{+}$	\\
$\Phi_{2}=\left(\begin{array}{c}
\phi_{2}^{+} \\ \frac{h_{2}+v_{2}+i\eta_{2}}{\sqrt{2}}	
\end{array}\right)$	&	${+1}/{3}^{-}$	&
$q^{2}_{L}=\left(\begin{array}{c}u^{2} \\ d^{2} \end{array}\right)_{L}$
	&	$0^{-}$	&
$\ell^{\mu}_{L}=\left(\begin{array}{c}\nu^{\mu} \\ e^{\mu} \end{array}\right)_{L}$
	&	$0^{+}$		\\
$\Phi_{3}=\left(\begin{array}{c}
\phi_{3}^{+} \\ \frac{h_{3}+v_{3}+i\eta_{3}}{\sqrt{2}}	
\end{array}\right)$	&	${+1}/{3}^{+}$	&
$q^{3}_{L}=\left(\begin{array}{c}u^{3} \\ d^{3} \end{array}\right)_{L}$
	&	$0^{+}$	&
$\ell^{\tau}_{L}=\left(\begin{array}{c}\nu^{\tau} \\ e^{\tau} \end{array}\right)_{L}$
	&	$-1^{+}$	\\   \hline\hline
	
\multicolumn{2}{c}{Scalar Singlets} &\multicolumn{4}{c}{SM Fermionic Singlets}	\\ \hline\hline
\begin{tabular}{c}$\chi  =\frac{\xi_{\chi}  +v_{\chi}  +i\zeta_{\chi}}{\sqrt{2}}$\\$\sigma$\end{tabular}	&
\begin{tabular}{c}${-1}/{3}^{+}$\\${-1}/{3}^{-}$\end{tabular}	&
\begin{tabular}{c}$u_{R}^{1,3}$\\$u_{R}^{2}$\\$d_{R}^{1,2,3}$\end{tabular}	&	 
\begin{tabular}{c}${+2}/{3}^{+}$\\${+2}/{3}^{-}$\\${-1}/{3}^{-}$\end{tabular}	&
\begin{tabular}{c}$e_{R}^{e}$\\$e_{R}^{\mu}$\\$e_{R}^{\tau}$\end{tabular}	&	
\begin{tabular}{c}${-4}/{3}^{+}$\\${-1}/{3}^{+}$\\${-4}/{3}^{-}$\end{tabular}\\   \hline \hline 

\multicolumn{2}{c}{Gauge bosons}&	\multicolumn{2}{c}{Non-SM Quarks}&	\multicolumn{2}{c}{Non-SM Leptons}\\ \hline \hline
\begin{tabular}{c}$W^{\pm}_{\mu}$\\$W^{3}_{\mu}$\end{tabular}	&
\begin{tabular}{c}$0^{+}$\\$0^{+}$\end{tabular}	&
\begin{tabular}{c}$\mathcal{T}_{L}$\\$\mathcal{T}_{R}$\end{tabular}	&
\begin{tabular}{c}${+1}/{3}^{-}$\\${+2}/{3}^{-}$\end{tabular}	&
\begin{tabular}{c}$\nu_{R}^{e,\mu,\tau}$\\$\mathcal{N}_{R}^{e,\mu,\tau}$\end{tabular} 	&	
\begin{tabular}{c}${+1}/{3}^{+}$\\$0^{+}$\end{tabular}\\
$B_{\mu}$	&	  $0^{+}$	&	
$\mathcal{J}^{1,2}_{L}$	&	  $0^{+}$	&
$\mathcal{E}_{L}^{1},\mathcal{E}_{R}^{2}$	&	$-1^{+}$	\\

$\Xi_{\mu}$	&	 $0^{+}$	&
$\mathcal{J}^{1,2}_{R}$	&	 ${-1}/{3}^{+}$	&
$\mathcal{E}_{R}^{1},\mathcal{E}_{L}^{2}$	&	${-2}/{3}^{+}$	\\ \hline \hline
\end{tabular}
\end{adjustbox}
\caption{Non-universal $X$ quantum number and $\mathbb{Z}_{2}$ parity for SM and non-SM fermions.}
\label{tab:Particle-content}
\end{table*}

The resulting Yukawa Lagrangian of the model for the neutral, charged lepton, up-like and down-like sectors are given by, respectively
\begin{equation}
\begin{split}
-\mathcal{L}_{N} &=
h_{3 \nu}^{e e}\overline{\ell^{e}_{L}}\tilde{\Phi}_{3}\nu^{e}_{R} + 
h_{3 \nu}^{e \mu}\overline{\ell^{e}_{L}}\tilde{\Phi}_{3}\nu^{\mu}_{R} + 
h_{3 \nu}^{e \tau}\overline{\ell^{e}_{L}}\tilde{\Phi}_{3}\nu^{\tau}_{R} +
h_{3 \nu}^{\mu e}\overline{\ell^{\mu}_{L}}\tilde{\Phi}_{3}\nu^{e}_{R} +
h_{3 \nu}^{\mu \mu}\overline{\ell^{\mu}_{L}}\tilde{\Phi}_{3}\nu^{\mu}_{R} + 
h_{3 \nu}^{\mu \tau}\overline{\ell^{\mu}_{L}}\tilde{\Phi}_{3}\nu^{\tau}_{R} \\ &+
g_{\chi \mathcal{N}}^{i j} \overline{\nu_{R}^{i\;C}} \chi^{*} \mathcal{N}_{R}^{j} +
\frac{1}{2} \overline{\mathcal{N}_{R}^{i\;C}} M^{ij}_{\mathcal{N}} \mathcal{N}_{R}^{j} + \mathrm{h.c.},
\end{split}
\label{eq:Neutrino-Lagrangian}
\end{equation}
\begin{equation}
\begin{split}
-\mathcal{L}_{E} &=
h_{3 e}^{e \mu}\overline{\ell^{e}_{L}}\Phi_{3}e^{\mu}_{R} + 
h_{3 e}^{\mu \mu}\overline{\ell^{\mu}_{L}}\Phi_{3}e^{\mu}_{R} + 
h_{3 e}^{\tau e}\overline{\ell^{\tau}_{L}}\Phi_{3}e^{e}_{R} +
h_{2 e}^{\tau \tau}\overline{\ell^{\tau}_{L}}\Phi_{2}e^{\tau}_{R} +
h_{1 E}^{e 1}\overline{\ell^{e}_{L}}\Phi_{1}\mathcal{E}_{R}^{1} + 
h_{1 \mathcal{E}}^{\mu 1}\overline{\ell^{\mu}_{L}}\Phi_{1}\mathcal{E}_{R}^{1} \\ &+
g_{\chi e}^{1 e}\overline{\mathcal{E}_{L}^{1}}\chi^{*} e^{e}_{R} + 
g_{\chi e}^{2 \mu}\overline{\mathcal{E}_{L}^{2}}\chi e^{\mu}_{R} +
g_{\chi \mathcal{E}}^{1}\overline{\mathcal{E}_{L}^{1}}\chi \mathcal{E}_{R}^{1}  + 
g_{\chi \mathcal{E}}^{2}\overline{\mathcal{E}_{L}^{2}}\chi^{*} \mathcal{E}_{R}^{2} + \mathrm{h.c.},
\end{split}
\label{eq:Electron-Lagrangian}
\end{equation}
\begin{equation}
\begin{split}
-\mathcal{L}_{U} &= 
h_{3 u}^{1 1}\overline{q_{L}^{1}}\tilde{\Phi}_{3}u_{R}^{1} + 
h_{2 u}^{1 2}\overline{q_{L}^{1}}\tilde{\Phi}_{2}u_{R}^{2} + 
h_{3 u}^{1 3}\overline{q_{L}^{1}}\tilde{\Phi}_{3}u_{R}^{3} + 
h_{1 u}^{2 2}\overline{q_{L}^{2}}\tilde{\Phi}_{1}u_{R}^{2} + 
h_{1 u}^{3 1}\overline{q_{L}^{3}}\tilde{\Phi}_{1}u_{R}^{1} + 
h_{1 u}^{3 3}\overline{q_{L}^{3}}\tilde{\Phi}_{1}u_{R}^{3} \\ &+
h_{2 \mathcal{T}}^{1} \overline{q_{L}^{1}}\tilde{\Phi}_{2}\mathcal{T}_{R} +
h_{1 \mathcal{T}}^{2} \overline{q_{L}^{2}}\tilde{\Phi}_{1}\mathcal{T}_{R} +
g_{\sigma u}^{1}\overline{\mathcal{T}_{L}}\sigma u_{R}^{1} + 
g_{\chi u}^{2}\overline{\mathcal{T}_{L}}\chi u_{R}^{2} \,+ 
g_{\sigma u}^{3}\overline{\mathcal{T}_{L}}\sigma u_{R}^{3} \,+ 
g_{\chi \mathcal{T}}\overline{\mathcal{T}_{L}}\chi \mathcal{T}_{R} + \mathrm{h.c.},
\end{split}
\label{eq:Up-Lagrangian}
\end{equation}
\begin{equation}
\begin{split}
-\mathcal{L}_{D} &= 
h_{1 \mathcal{J}}^{1 1}\overline{q_{L}^{1}}{\Phi}_{1}\mathcal{J}_{R}^{1} + 
h_{2 \mathcal{J}}^{2 1}\overline{q_{L}^{2}}{\Phi}_{2}\mathcal{J}_{R}^{1} + 
h_{3 \mathcal{J}}^{3 1}\overline{q_{L}^{3}}{\Phi}_{3}\mathcal{J}_{R}^{1} +
h_{1 \mathcal{J}}^{1 2}\overline{q_{L}^{1}}{\Phi}_{1}\mathcal{J}_{R}^{2} + 
h_{2 \mathcal{J}}^{2 2}\overline{q_{L}^{2}}{\Phi}_{2}\mathcal{J}_{R}^{2} + 
h_{3 \mathcal{J}}^{3 2}\overline{q_{L}^{3}}{\Phi}_{3}\mathcal{J}_{R}^{2} \\ &+ 
h_{3 d}^{2 1}\overline{q_{L}^{2}}{\Phi}_{3}d_{R}^{1} +
h_{3 d}^{2 2}\overline{q_{L}^{2}}{\Phi}_{3}d_{R}^{2} +
h_{3 d}^{2 3}\overline{q_{L}^{2}}{\Phi}_{3}d_{R}^{3} +
h_{2 d}^{3 1}\overline{q_{L}^{3}}{\Phi}_{2}d_{R}^{1} +
h_{2 d}^{3 2}\overline{q_{L}^{3}}{\Phi}_{2}d_{R}^{2} +
h_{2 d}^{3 3}\overline{q_{L}^{3}}{\Phi}_{2}d_{R}^{3} \\ &+ 
g_{\sigma d}^{1 1}\overline{\mathcal{J}_{L}^{1}}\sigma^{*} d_{R}^{1} + 
g_{\sigma d}^{1 1}\overline{\mathcal{J}_{L}^{1}}\sigma^{*} d_{R}^{2} +
g_{\sigma d}^{1 3}\overline{\mathcal{J}_{L}^{1}}\sigma^{*} d_{R}^{3} + 
g_{\sigma d}^{2 1}\overline{\mathcal{J}_{L}^{2}}\sigma^{*} d_{R}^{1} +
g_{\sigma d}^{2 2}\overline{\mathcal{J}_{L}^{2}}\sigma^{*} d_{R}^{2} +
g_{\sigma d}^{2 3}\overline{\mathcal{J}_{L}^{2}}\sigma^{*} d_{R}^{3} \\ &+ 
g_{\chi \mathcal{J}}^{1}\overline{\mathcal{J}_{L}^{1}}\chi^{*} \mathcal{J}_{R}^{1} + 
g_{\chi \mathcal{J}}^{2}\overline{\mathcal{J}_{L}^{2}}\chi^{*} \mathcal{J}_{R}^{2} + \mathrm{h.c.},
\end{split}
\label{eq:Down-Lagrangian}
\end{equation}
where $\widetilde{\Phi}=i\sigma_2 \Phi^*$ are the scalar doublet conjugates. 


\section{Mass matrices}
The up-like quark sector is described in the bases $\mathbf{U}=(u^{1},u^{2},u^{3},\mathcal{T})$ and $\mathbf{u}=(u, c, t, T)$, where the former is the flavor basis while the latter is the mass basis. The mass term in the flavor basis turns out to be
\begin{equation}
-\mathcal{L}_{U} = \overline{\mathbf{U}_{L}} \mathbb{M}_{U} \mathbf{U}_{R} + \mathrm{h.c.},
\end{equation}
where $\mathbb{M}_{U}$ is
\begin{equation}
\mathbb{M}_{U} = \frac{1}{\sqrt{2}}
\begin{pmatrix}
h_{3 u}^{1 1}v_{3}	&	h_{2 u}^{1 2}v_{2}	&h_{3 u}^{1 3}v_{3}	&h_{2 \mathcal{T}}^{1}v_{2}	\\
0	&	h_{1 u}^{2 2}v_{1}	&	0	&h_{1 \mathcal{T}}^{2}v_{1}	\\
h_{1 u}^{3 1}v_{1}	&	0	&	h_{1 u}^{3 3}v_{1}	&	0	\\
0	&	g_{\chi u}^{2}v_{\chi}	&	0	&	g_{\chi \mathcal{T}}v_{\chi}
\end{pmatrix}.
\end{equation}
Since the determinant of $\mathbb{M}_{U}$ is non-vanishing, the four up-like quarks acquire masses. The four mass eigenvalues are
\begin{equation}
\label{eq:Up-Quarks-masses}
\begin{split}
m_{u}^{2}&=\frac{\left(h_{3 u}^{1 1}h_{1 u}^{3 3}-h_{3 u}^{1 3}h_{1 u}^{3 1}\right)^{2}}{(h_{1 u}^{3 3})^{2}+(h_{1 u}^{3 1})^{2}}
\frac{v_{3}^{2}}{2},	\qquad 
m_{c}^{2} =\frac{\left(h_{1 u}^{2 2}g_{\chi \mathcal{T}}-h_{1 \mathcal{T}}^{2}g_{\chi u}^{2}\right)^{2}}
{(g_{\chi \mathcal{T}})^{2}+(g_{\chi u}^{2})^{2}}\frac{v_{1}^{2}}{2},	\\
m_{t}^{2}&=\left[(h_{1 u}^{3 3})^{2}+(h_{1 u}^{3 1})^{2} \right]\frac{v_{1}^{2}}{2},\qquad
m_{T}^{2} = \left[(g_{\chi \mathcal{T}})^{2}+(g_{\chi u}^{2})^{2} \right]\frac{v_{\chi}^{2}}{2},
\end{split}
\end{equation}
where the VH has been employed. The heavy quarks $T$ and $t$ acquire masses through $v_{\chi}$ and $v_{1}$, respectively. The $c$ quark acquire mass also through $v_{1}$, however, this exhibits the difference of the Yukawa coupling constants because of the see-saw with the exotic quark $T$. Finally, the $u$ quark acquire mass through $v_{3}$ with the same suppression mechanisms of $c$ quark but with $t$ instead of $T$. 

The down-like quarks are described in the bases $\mathbf{D}=(d^{1},d^{2},d^{3},\mathcal{J}^{1},\mathcal{J}^{2})$ and $\mathbf{d}=(d, s, b, J^{1}, J^{2})$, where the former is the flavor basis while the latter is the mass basis. The matrix in the flavor basis is
\begin{equation}
-\mathcal{L}_{D} = \overline{\mathbf{D}_{L}} \mathbb{M}_{D} \mathbf{D}_{R} + \mathrm{h.c.},
\end{equation}
where $\mathbb{M}_{D}$ turns out to be 
\begin{equation}
\label{eq:Down-mass-matrix}
\mathbb{M}_{D} = \frac{1}{\sqrt{2}}
\begin{pmatrix}
0	&	0	&	0	&	h_{1 \mathcal{J}}^{1 1}v_{1}	&	h_{1 \mathcal{J}}^{1 2}v_{1}	\\
h_{3 d}^{2 1}v_{3}	&	h_{3 d}^{2 2}v_{3}	&	h_{3 d}^{2 3}v_{3}	&	h_{2 \mathcal{J}}^{2 1}v_{2}	&	h_{2 \mathcal{J}}^{2 2}v_{2}	\\
h_{2 d}^{3 1}v_{2}	&	h_{2 d}^{3 2}v_{2}	&	h_{2 d}^{3 3}v_{2}	&	h_{3 \mathcal{J}}^{3 1}v_{3}	&	h_{3 \mathcal{J}}^{3 2}v_{3}	\\
0	&	0	&	0	&	g_{\chi \mathcal{J}}^{1}v_{\chi}	&0\\
0	&	0	&	0	&	0	&	g_{\chi \mathcal{J}}^{2}v_{\chi}
\end{pmatrix}
\end{equation}

Although $\mathbb{M}_{D}$ has a vanishing determinant and the quark $d$ remains massless, it can acquire a small mass with radiative corrections, 
given by the expression
\begin{equation}
\begin{split}
\Sigma_{d}^{1k} &= \sum_{i=1,2} \frac{f_{\sigma}{g_{\sigma d}^{i 1}}^{*}{h_{k \mathcal{J}}^{k i}}v_{k}}{(4\pi)^{2}m_{Ji}}
C_{0}\left( \frac{m_{\sigma}}{m_{Ji}},\frac{m_{hk}}{m_{Ji}} \right)
\end{split}
\end{equation}
where $k=1,2,3$, $f_{\sigma}$ is the trilinear coupling constant involving $\sigma$ and $\Phi_{1,2,3}$, and $C_{0}\left( x,y \right)$ is\cite{martinez-rad-corr-331}
\begin{equation}
\begin{split}
C_{0}\left( x,y \right) &= \frac{1}{(1-x^{2})(1-y^{2})(x^{2}-y^{2})} 
\left\lbrace x^{2}y^{2}\ln\left(\frac{x^{2}}{y^{2}} \right) 
- x^{2}\ln x^{2} + y^{2}\ln y^{2} \right\rbrace 
\end{split}
\end{equation}
Thus, up to one-loop correction the mass matrix is
\begin{equation}
\mathbb{M}_{D} = \frac{1}{\sqrt{2}}
\begin{pmatrix}
\Sigma_{d}^{11}	&	\Sigma_{d}^{12}	&	\Sigma_{d}^{13}	&	h_{1 \mathcal{J}}^{1 1}v_{1}	&	h_{1 \mathcal{J}}^{1 2}v_{1}	\\
h_{3 d}^{2 1}v_{3}	&	h_{3 d}^{2 2}v_{3}	&	h_{3 d}^{2 3}v_{3}	&	h_{2 \mathcal{J}}^{2 1}v_{2}	&	h_{2 \mathcal{J}}^{2 2}v_{2}	\\
h_{2 d}^{3 1}v_{2}	&	h_{2 d}^{3 2}v_{2}	&	h_{2 d}^{3 3}v_{2}	&	h_{3 \mathcal{J}}^{3 1}v_{3}	&	h_{3 \mathcal{J}}^{3 2}v_{3}	\\
0	&	0	&	0	&	g_{\chi \mathcal{J}}^{1}v_{\chi}	&0\\
0	&	0	&	0	&	0	&	g_{\chi \mathcal{J}}^{2}v_{\chi}
\end{pmatrix}	
\end{equation}
whose determinant does not vanish. The eigenvalues are, according to VH, 
\begin{equation}
\label{eq:Down-Quarks-masses-Light}
\begin{split}
m_{d}^{2} &= \frac{\left[
\left(\Sigma_{d}^{11}h_{3 d}^{2 2}-\Sigma_{d}^{12}h_{3 d}^{2 1} \right)h_{2 d}^{3 3} + 
\left(\Sigma_{d}^{13}h_{3 d}^{2 1}-\Sigma_{d}^{11}h_{3 d}^{2 3} \right)h_{2 d}^{3 2} + 
\left(\Sigma_{d}^{12}h_{3 d}^{2 3}-\Sigma_{d}^{13}h_{3 d}^{2 2} \right)h_{2 d}^{3 1}
\right]^2}{
\left[(h_{3 d}^{2 1})^2+(h_{3 d}^{2 2})^2 \right](h_{2 d}^{3 3})^2 + 
\left[(h_{3 d}^{2 3})^2+(h_{3 d}^{2 1})^2 \right](h_{2 d}^{3 2})^2 + 
\left[(h_{3 d}^{2 2})^2+(h_{3 d}^{2 3})^2 \right](h_{2 d}^{3 1})^2},	\\
m_{s}^{2} &= \frac{
\left[(h_{3 d}^{2 1})^2+(h_{3 d}^{2 2})^2 \right](h_{2 d}^{3 3})^2 + 
\left[(h_{3 d}^{2 3})^2+(h_{3 d}^{2 1})^2 \right](h_{2 d}^{3 2})^2 + 
\left[(h_{3 d}^{2 2})^2+(h_{3 d}^{2 3})^2 \right](h_{2 d}^{3 1})^2}
{(h_{2 d}^{3 3})^2+(h_{2 d}^{3 2})^2+(h_{2 d}^{3 1})^2}\frac{v_{3}^{2}}{2},
\end{split}
\end{equation}
while the masses of $b$, $J^{1}$ and $J^{2}$ are
\begin{equation}
\label{eq:Down-Quarks-masses-Heavy}
\begin{split}
m_{b}^{2}  = \left[(h_{2 d}^{3 3})^2+(h_{2 d}^{3 2})^2+(h_{2 d}^{3 1})^2 \right]\frac{v_{2}^{2}}{2},\quad
m_{J1}^{2} = (g_{\chi \mathcal{J}}^{1})^2\frac{v_{\chi}^{2}}{2},	\quad 
m_{J2}^{2} = (g_{\chi \mathcal{J}}^{2})^2\frac{v_{\chi}^{2}}{2}.
\end{split}
\end{equation}
The heaviest quarks $J^{1}$ and $J^{2}$ acquire masses at TeV scale with $v_{\chi}$, while $b$ quark get mass through $v_{2}$ at GeV. The strange quark acquire mass with $v_{3}$ at hundreds of MeV, and the lightest $d$ did not acquire mass at tree-level but at one-loop, where radiative corrections supress its mass. 

The neutrinos involve both Dirac and Majorana masses in their Yukawa Lagrangian. 
The flavor and mass basis are $\mathbf{N}_{L}=(\nu^{e,\mu,\tau}_{L},{\nu^{e,\mu,\tau}_{R}}^{C},{\mathcal{N}^{e,\mu,\tau}_{R}}^{C})$ and $\mathbf{n}_{L}=(\nu^{1,2,3}_{L},N^{1,2,3}_{L},\tilde{N}^{1,2,3}_{L})$, respectively. The mass term expressed in the flavor basis is
\begin{equation}
-\mathcal{L}_{N} = \frac{1}{2} \overline{\mathbf{N}_{L}^{C}} \mathbb{M}_{N} \mathbf{N}_{L},
\end{equation}
where the mass matrix has the following block structure
\begin{equation}
\mathbb{M}_{N} = 
\left(\begin{array}{c c c}
0	&	\mathcal{M}_{\nu}^{\mathrm{T}}	&	0	\\
\mathcal{M}_{\nu}	&	0	&	\mathcal{M}_{\mathcal{N}}^{\mathrm{T}}	\\
0	&	\mathcal{M}_{\mathcal{N}}	&	M_{\mathcal{N}}
\end{array}\right),
\end{equation}
with $\mathcal{M}_{\mathcal{N}} = \mathrm{diag}\left( \begin{matrix}
h_{\mathcal{N}}^{1},h_{\mathcal{N}}^{2},h_{\mathcal{N}}^{3}
\end{matrix} \right)\frac{v_{\chi}}{\sqrt{2}}$ the Dirac mass in the ($\nu_{R}^{C}$, $\mathcal{N}_{R}$) basis, and
\begin{equation}
\label{eq:m_nu_original_parameters}
\mathcal{M}_{\nu} = \frac{v_{3}}{\sqrt{2}}\left(\begin{matrix}
h_{3 \nu}^{e e}	&	h_{3 \nu}^{e \mu}	&	h_{3 \nu}^{e \tau}	\\
h_{3 \nu}^{\mu e}&	h_{3 \nu}^{\mu \mu}	&	h_{3 \nu}^{\mu \tau}	\\
0	&	0	&	0	\end{matrix}\right),
\end{equation}
is a Dirac mass matrix for ($\nu_{L}$, $\nu_{R}$). $M_{\mathcal{N}} = \mu_{\mathcal{N}} \mathbb{I}_{3\times 3}$ is the Majorana mass of $\mathcal{N}_{R}$.
The ISS, together with the VH $v_{\chi}\gg v_{3}\gg |M_{\mathcal{N}}|$ yields
\begin{equation}
\left(\mathbb{V}_{L,\mathrm{SS}}^{N}\right)^{\dagger}\mathbb{M}_{N} \mathbb{V}_{L,\mathrm{SS}}^{N} = 
\begin{pmatrix}
m_{\nu}	&	0	&	0	\\
0	&	m_{N}	&	0	\\
0	&	0	&	m_{\tilde{N}}
\end{pmatrix}
\end{equation}
where the resultant $3\times 3$ blocks are\cite{catano2012}
\begin{equation}
\label{eq:Neutrino-block-mass-matrices}
\begin{split}
m_{\nu} =	\mathcal{M}_{\nu}^{\mathrm{T}} \left( \mathcal{M}_{\mathcal{N}} \right)^{-1} M_{\mathcal{N}} \left( \mathcal{M}_{\mathcal{N}}^{\mathrm{T}} \right)^{-1} \mathcal{M}_{\nu}, \quad 
M_{N} \approx \mathcal{M}_{\mathcal{N}}-{M}_{\mathcal{N}},	\quad	
M_{\tilde{N}}\approx \mathcal{M}_{\mathcal{N}}+{M}_{\mathcal{N}}.
\end{split}
\end{equation}


The charged leptons are described in the bases of flavor $\mathbf{E}=(e^{e},e^{\mu} , e^{\tau}, \mathcal{E}^{1}, \mathcal{E}^{2})$ and mass $\mathbf{e}=(e, \mu, \tau, E^{1}, E^{2})$. The mass term obtained from the Yukawa Lagrangian is
\begin{equation}
-\mathcal{L}_{E} = \overline{\mathbf{E}_{L}}\mathbb{M}_{E} \mathbf{E}_{R} + \mathrm{h.c.}
\end{equation}
where $\mathbb{M}_{E}$ turns out to be 
\begin{equation}
\label{eq:Electron-mass-matrix}
\begin{split}
\mathbb{M}_{E} =  \frac{1}{\sqrt{2}}
\begin{pmatrix}
0	&	h_{3 e}^{e \mu}v_{3}	&	0	&	h_{1 \mathcal{E}}^{e 1}v_{1}	&	0	\\
0	&	h_{3 e}^{\mu \mu}v_{3}	&	0	&	h_{1 \mathcal{E}}^{\mu 1}v_{1}	&	0	\\
h_{3 e}^{\tau e}v_{3}	&	0	&	h_{2 e}^{\tau \tau}v_{2}	&	0	&	0	\\
g_{\chi e}^{1 e}v_{\chi}	&	0	&	0	&	g_{\chi \mathcal{E}}^{1}v_{\chi}	&0\\
0	&	g_{\chi e}^{2 \mu}v_{\chi}	&	0	&0	&	g_{\chi \mathcal{E}}^{2}v_{\chi}
\end{pmatrix}
\end{split}
\end{equation}

The determinant of $\mathbb{M}_{E}$ is non-vanishing ensuring that the five charged leptons acquire masses. By using the VH, the eigenvalues are
\begin{equation}
\label{eq:Charged-Lepton-masses}
\begin{split}
m_{e}^{2} &= \frac{\left(h_{3 e}^{e \mu}h_{1 \mathcal{E}}^{\mu 1}-h_{3 e}^{\mu \mu}h_{1 \mathcal{E}}^{e 1}\right)^2}
{(h_{1 \mathcal{E}}^{e 1})^2+(h_{1 \mathcal{E}}^{\mu 1})^2}\frac{v_{3}^{2}}{2},		\qquad
m_{\mu}^{2}  = \frac{\left(h_{3 e}^{e \mu}h_{1 \mathcal{E}}^{e 1}+h_{3 e}^{\mu \mu}h_{1 \mathcal{E}}^{\mu 1}\right)^2}
{(h_{1 \mathcal{E}}^{e 1})^2+(h_{1 \mathcal{E}}^{\mu 1})^2}\frac{v_{3}^{2}}{2}
+\frac{\left(h_{3 e}^{\tau e}\right)^{2} v_{3}^{2}}{2},	\\
m_{\tau}^{2} &= \frac{\left(h_{2 e}^{\tau \tau}\right)^{2}v_{2}^{2}}{2},	\qquad 
m_{E{1}}^{2}  = \left[\left(g_{\chi \mathcal{E}}^{1}\right)^{2}+\left(g_{\chi e}^{1 e}\right)^{2}\right]\frac{v_{\chi}^{2}}{2},	\qquad 
m_{E{2}}^{2}  = \left[\left(g_{\chi \mathcal{E}}^{2}\right)^2+\left(g_{\chi e}^{2\mu}\right)^2\right]\frac{v_{\chi}^{2}}{2}.
\end{split}
\end{equation}
The exotic $E^{1}$ and $E^{2}$ leptons acquired mass at the TeV scale. $\tau$ got mass at the GeV scale with $v_{2}$. $\mu$ and $e$ have acquired mass through $v_{3}$, the smallest VEV at hundreds of MeV. Moreover, the $e$ mass is suppressed by the difference between the Yukawa coupling constants, which can be assumed to be at the same order of magnitude. Finally, because the mixing angle between $e$ and $\mu$ is not small and can contribute importantly to the PMNS matrix, is set as a free parameter named $\theta_{12}^{E}$.

\section{PMNS fitting}
\label{PMNS-fitting}
By replacing the Dirac mass matrix from \eqref{eq:m_nu_original_parameters} into the light mass eigenvalues in \eqref{eq:Neutrino-block-mass-matrices} and redefining the Yukawa coupling constants by a polar parametrization
\begin{equation}
\label{eq:m_nu_polar_parameters}
\mathcal{M}_{\nu} = \frac{v_{3}}{\sqrt{2}{\rho}}\left(\begin{matrix}
{\rho}h_{\nu}^{e}c_{\nu}^{e}&	{\rho}h_{\nu}^{\mu}c_{\nu}^{\mu}	&	{\rho}h_{\nu}^{\tau}c_{\nu}^{\tau}	\\
{h_{\nu}^{e}s_{e}}	&	{h_{\nu}^{\mu}s_{\mu}}	&	{h_{\nu}^{\tau}s_{\tau}}	\\
0	&	0	&	0	\end{matrix}\right),
\end{equation}
the effective mass matrix for the light neutrinos is
\begin{equation}
\label{eq:Neutrino-mass-matrix_polar_parameters}
m_{\nu} = \frac{\mu_{\mathcal{N}} v_{3}^{2}}{{\left(h_{\mathcal{N}}^{1}\right)}^{2}v_{\chi}^{2}}
\left( \begin{matrix}
	\left({h_{\nu}^{e}}\right)^{2}	&h_{\nu}^{e}h_{\nu}^{\mu}c_{\nu}^{e\mu} 	&h_{\nu}^{e}h_{\nu}^{\tau}c_{\nu}^{e\tau}\\
	h_{\nu}^{e}h_{\nu}^{\mu}c_{\nu}^{e\mu}	&\left({h_{\nu}^{\mu}}\right)^{2}	&h_{\nu}^{\mu}h_{\nu}^{\tau}c_{\nu}^{\mu\tau}\\
	h_{\nu}^{e}h_{\nu}^{\tau}c_{\nu}^{e\tau}	&h_{\nu}^{\mu}h_{\nu}^{\tau}c_{\nu}^{\mu\tau}&\left({h_{\nu}^{\tau}}\right)^{2}
\end{matrix} \right),
\end{equation}
where $c_{\nu}^{\alpha\beta}=\cos(\theta_{\nu}^{\alpha}-\theta_{\nu}^{\beta})$. While the mass scale is fixed by the constrain
${\mu_{\mathcal{N}} v_{3}^{2}}/{{\left(h_{\mathcal{N}}^{1}v_{\chi}\right)}^{2}} = 50\mathrm{\,meV}$, the other parameters can be explored with Montecarlo procedures in order to reproduce neutrino oscillation data\cite{neutrinodata}. Since $m_{\nu}$ only depends on angle differences, $\theta_{\nu}^{e}$ is set null. Tab. \ref{tab:Neutrino-parameters} presents some domains were neutrino oscillation data are reproduced at 3$\sigma$\cite{neutrinodata} in function of the charged lepton mixing angle $\theta_{12}^{E}$. Thus, the model is able to reproduce neutrino oscillation data in NO and IO schemes. 

\begin{table*}
\centering
\begin{adjustbox}{width=0.9\textwidth,center}
\begin{tabular}{c||ccc|cc}
$\theta_{12}^{E}$	&$h_{\nu}^{e}$	&$h_{\nu}^{\mu}$	&$h_{\nu}^{\tau}$	&$\theta_{\nu}^{\mu}$	&$\theta_{\nu}^{\tau}-\theta_{\nu}^{\mu}$\\\hline\hline
\multicolumn{6}{c}{Normal Ordering}\\ \hline\hline
$0^{\mathrm{o}}$	&	
\begin{tabular}{c}
$0.270\pm0.007$
\end{tabular}	&					
\begin{tabular}{c}					
$0.738\pm0.040$
\end{tabular}	&					
\begin{tabular}{c}						
$0.747\pm0.040$
\end{tabular}	&						
\begin{tabular}{c}						
$\pm(39.49\pm 2.99)$
\end{tabular}	&					
\begin{tabular}{c}	
$\pm(38.79\pm 0.78)$
\end{tabular}\\\hline
$15^{\mathrm{o}}$	&
\begin{tabular}{c}
$0.294\pm0.008$
\end{tabular}	&					
\begin{tabular}{c}					
$0.737\pm0.045$
\end{tabular}	&					
\begin{tabular}{c}						
$0.738\pm0.043$
\end{tabular}	&						
\begin{tabular}{c}						
$\pm(66.73\pm 1.02)$
\end{tabular}	&					
\begin{tabular}{c}					
$\pm(33.32\pm 0.81)$
\end{tabular}\\\hline
$30^{\mathrm{o}}$	&	
\begin{tabular}{c}
$0.400\pm0.008$
\end{tabular}	&					
\begin{tabular}{c}					
$0.689\pm0.035$
\end{tabular}	&					
\begin{tabular}{c}						
$0.734\pm0.035$
\end{tabular}	&						
\begin{tabular}{c}			
$\pm(46.38\pm 1.91)$
\end{tabular}	&					
\begin{tabular}{c}					
$\pm(27.53\pm 1.12)$
\end{tabular}\\\hline
$45^{\mathrm{o}}$	&	
\begin{tabular}{c}
$0.495\pm0.003$
\end{tabular}	&					
\begin{tabular}{c}					
$0.548\pm0.004$
\end{tabular}	&					
\begin{tabular}{c}						
$0.796\pm0.005$
\end{tabular}	&
\begin{tabular}{c}
$\pm(42.61\pm 0.82)$
\end{tabular}	&
\begin{tabular}{c}
$\pm(19.10\pm 0.75)$
\end{tabular}\\\hline
\multicolumn{6}{c}{Inverted Ordering}\\ \hline\hline
$0^{\mathrm{o}}$	&	
\begin{tabular}{c}
$0.984\pm0.006$
\end{tabular}	&					
\begin{tabular}{c}					
$0.725\pm0.031$
\end{tabular}	&					
\begin{tabular}{c}						
$0.700\pm0.032$
\end{tabular}	&						
\begin{tabular}{c}						
$\pm(81.88\pm 0.84)$
\end{tabular}	&					
\begin{tabular}{c}					
$\mp(163.17\pm0.56)$
\end{tabular}\\\hline
$1^{\mathrm{o}}$	&	
\begin{tabular}{c}
$0.982\pm0.006$
\end{tabular}	&					
\begin{tabular}{c}					
$0.732\pm0.030$
\end{tabular}	&					
\begin{tabular}{c}						
$0.695\pm0.031$
\end{tabular}	&						
\begin{tabular}{c}						
$\pm(81.57\pm 0.74)$
\end{tabular}	&					
\begin{tabular}{c}					
$\pm(161.91\pm0.55)$
\end{tabular}\\\hline
$2^{\mathrm{o}}$	&	
\begin{tabular}{c}
$0.980\pm0.006$
\end{tabular}	&					
\begin{tabular}{c}					
$0.747\pm0.022$
\end{tabular}	&					
\begin{tabular}{c}						
$0.681\pm0.022$
\end{tabular}	&						
\begin{tabular}{c}						
$\pm(81.54\pm 0.51)$
\end{tabular}	&					
\begin{tabular}{c}
$\mp(160.77\pm0.50)$
\end{tabular}\\\hline
$3^{\mathrm{o}}$	&	
\begin{tabular}{c}
$0.978\pm0.006$
\end{tabular}	&					
\begin{tabular}{c}					
$0.759\pm0.014$
\end{tabular}	&					
\begin{tabular}{c}						
$0.671\pm0.013$
\end{tabular}	&						
\begin{tabular}{c}						
$\pm(81.51\pm 0.32)$
\end{tabular}	&					
\begin{tabular}{c}					
$\mp(159.56\pm0.46)$
\end{tabular}\\\hline
\end{tabular}
\end{adjustbox}
\caption{Some parameter domains which reproduce data for NO and IO from \cite{neutrinodata}. $\theta_{\nu}^{e}=0$.}
\label{tab:Neutrino-parameters}
\end{table*}

\section{Discussion and Conclusions}
The model proposes an extension to the SM by including a new nonuniversal abelian interaction $\mathrm{U(1)}_{X}$ and a discrete symmetry $\mathbb{Z}_{2}$, with extended fermion and scalar sectors such that chiral anomalies get cancelled and the majority of fermions can acquire mass. The existence of the VH, together with the suited Yukawa coupling constants yield mass matrices whose eigenvalues suggest the presence of suppression mechanisms which could offer a fermionic spectrum spanning different order of magnitude of mass, from units of MeV to units of TeV. Such mass eigenvalues are outlined in tab. \ref{tab:Fermion-masses}. 

\begin{table}[h]
\centering
\begin{adjustbox}{width=7cm,center}
\begin{tabular}{c cc cc}
	&	\multicolumn{2}{c}{Leptons}	&	\multicolumn{2}{c}{Quarks}\\ \hline\hline
Family	&	&	Mass	& 	&	Mass	\\ \hline\hline
1&$\nu_{L}^{1}$	&	$\dfrac{\mu_{\mathcal{N}} v_{3}^{2}}{{\left(h_{\mathcal{N}}^{1}\right)}^{2}v_{\chi}^{2}}h_{\nu1}^{2}$	&
$u$	&	$\dfrac{h_{u}^{2}-{h_{u}'}^{2}}{h_{t}}\dfrac{v_{3}}{\sqrt{2}} $\\
2&$\nu_{L}^{2}$	&	$\dfrac{\mu_{\mathcal{N}} v_{3}^{2}}{{\left(h_{\mathcal{N}}^{1}\right)}^{2}v_{\chi}^{2}}h_{\nu2}^{2}$	&
$c$	&	$\dfrac{h_{c}^{2}-{h_{c}'}^{2}}{h_{T}}\dfrac{v_{1}}{\sqrt{2}} $\\
3&$\nu_{L}^{3}$	&	$\dfrac{\mu_{\mathcal{N}} v_{3}^{2}}{{\left(h_{\mathcal{N}}^{1}\right)}^{2}v_{\chi}^{2}}h_{\nu3}^{2}$	&
$t$	&	$\dfrac{h_{t}v_{1}}{\sqrt{2}}$\\
Exot&\begin{tabular}{c}$N_{L}^{i}$ \end{tabular}	&	
\begin{tabular}{c}
$\dfrac{h_{\mathcal{N}}^{i}v_{\chi}}{\sqrt{2}}\mp\mu_{\mathcal{N}}$
\end{tabular}	&	
$T$	&	$\dfrac{h_{T}v_{\chi}}{\sqrt{2}}$\\	\hline
1&$e$	&	$\dfrac{h_{\ell}^{2}-{h_{\ell}'}^{2}}{h_{v1}}\dfrac{v_{3}}{\sqrt{2}} $	&
$d$	&	$\dfrac{\Sigma_{d}h_{d}^{2}}{h_{s}^{2}+{h_{s}'}^{2}}$\\
2&$\mu$	&	$\dfrac{h_{\ell}^{2}+{h_{\ell}'}^{2}}{h_{v1}}\dfrac{v_{3}}{\sqrt{2}} $	&
$s$	&	$\dfrac{h_{s}^{2}+{h_{s}'}^{2}}{h_{b}}\dfrac{v_{3}}{\sqrt{2}} $\\
3&$\tau$&	$\dfrac{h_{\tau}v_{2}}{\sqrt{2}}$	&
$b$	&	$\dfrac{h_{b}v_{2}}{\sqrt{2}}$\\
Exot&$E^{1}$&	$\dfrac{h_{E1} v_{\chi}}{\sqrt{2}}$	&
$J^{1}$	&	$\dfrac{h_{J1}v_{\chi}}{\sqrt{2}}$\\
Exot&$E^{2}$&	$\dfrac{h_{E2} v_{\chi}}{\sqrt{2}}$	&
$J^{2}$	&	$\dfrac{h_{J2}v_{\chi}}{\sqrt{2}}$\\	\hline
\end{tabular}
\end{adjustbox}
\caption{Summary of fermion masses showing their VEVs as well as the suppression mechanism if it is involved. The orders of magnitude of $v_{\chi}$, $v_{1}$, $v_{2}$, $v_{3}$ and $\mu_{\mathcal{N}}$ are units of TeV, hundreds of GeV, units of GeV, hundreds of MeV and units of MeV, respectively.}
\label{tab:Fermion-masses}
\end{table}

These eigenvalues are produced by the very structure of the mass matrices where a heavier mass suppress a lighter one. For instance, the subblock involving the $u$ and $t$ quarks is
\begin{equation*}
\mathbb{M}_{ut} \propto
\begin{pmatrix}
h_{3 u}^{1 1}v_{3}	&\left| \right. & h_{3 u}^{1 3}v_{3}	\\
\text{\textemdash}&\text{\textemdash}&\text{\textemdash}\\
h_{1 u}^{3 1}v_{1}	&\left| \right. & h_{1 u}^{3 3}v_{1}
\end{pmatrix},
\end{equation*}
whose eigenvalues turn out to be (with the assumption $v_{3}/v_{1}\ll 1$ from VH)
\begin{equation*}
\begin{split}
m_{u}^{2}&=\frac{\left(h_{3 u}^{1 1}h_{1 u}^{3 3}-h_{3 u}^{1 3}h_{1 u}^{3 1}\right)^{2}}{(h_{1 u}^{3 3})^{2}+(h_{1 u}^{3 1})^{2}}\frac{v_{3}^{2}}{2},	\qquad
m_{t}^{2} =\left[(h_{1 u}^{3 3})^{2}+(h_{1 u}^{3 1})^{2} \right]\frac{v_{1}^{2}}{2}+\frac{\left(h_{3 u}^{1 1}h_{1 u}^{3 3}+h_{3 u}^{1 3}h_{1 u}^{3 1}\right)^{2}}{(h_{1 u}^{3 3})^{2}+(h_{1 u}^{3 1})^{2}}\frac{v_{3}^{2}}{2}. 
\end{split}
\end{equation*}
This result is repeated with the $c$ quark and the $e$. Thus, the FMH is addressed from the structure of the mass matrices, which is also produced by the nonuniversal interaction $\mathrm{U(1)}_{X}$ and the discrete symmetry $\mathbb{Z}_{2}$. In contrast, the $d$ quark acquires mass through radiative corrections involving $\sigma$, $\Phi_{1,2,3}$ and the exotic quarks $\mathcal{J}^{1,2}$. 

On the other hand, the existence of parameter domains where neutrino oscillation data can be repoduced outlines the suitability of the model to understand neutrino masses from the nonuniversal interaction and the ISS with the right-handed neutrinos $\nu_{R}^{e,\mu,\tau}$ and the Majorana fermions $\mathcal{N}^{1,2,3}$. Moreover, the model is able to reproduce NO and IO schemes according to current neutrino oscillation data. The squared mass differences and PMNS mixing angles were obtained without fine-tunings. 

\subsubsection*{Acknowledgment}
This work was supported by COLCIENCIAS in Colombia.

\end{document}